\documentclass[preprints,article,accept,pdftex,moreauthors]{Definitions/mdpi} 
\usepackage[normalem]{ulem}
\newif\ifrev
\revtrue 
\ifrev

\else

\fi
\usepackage[labelformat=simple]{subcaption}

\DeclareCaptionLabelFormat{subcaptionlabel}{\normalfont(\textbf{#2}\normalfont)}
\firstpage{1} 
\pubvolume{1}
\issuenum{1}
\articlenumber{0}
\pubyear{2022}
\copyrightyear{2022}
\datereceived{28 October 2022} 
\dateaccepted{1 December 2022} 
\datepublished{} 
\hreflink{https://doi.org/} 
\pdfoutput=1

\usepackage[none]{hyphenat}
\usepackage{makecell}
\usepackage{subcaption}
\usepackage{url}
\usepackage{placeins}

\Title{Human Exposure to Non-Ionizing Radiation from Indoor Distributed Antenna System: Shopping Mall Measurement
~Analysis}

\TitleCitation{Human Exposure to Non-Ionizing Radiation from Indoor Distributed Antenna System: Shopping Mall Measurement
 Analysis}

\Author{Júlia da L. A. Silva
$^{1}$ 
\orcidA{}, 
Vicente A. de Sousa, Jr. $^{1,*}$\orcidB{},
Marcio E. C. Rodrigues~$^{2}$\orcidC{}, 
\mbox{Fred Sizenando Rossiter Pinheiro~$^{2}$\orcidD{}, }
Gutembergue Soares da Silva~$^{2}$\orcidE{},
Halysson B. Mendonça~$^{3}$\orcidF{},
\mbox{Ricardo Q. de F. H. Silva~$^{4}$\orcidG{}, }
João V. L. da Silva $^{4}$\orcidH{}, 
Fernanda E. S. Galdino $^{2}$\orcidI{},
Vitor F. C. de Carvalho $^{4}$\orcidJ{} 
\mbox{and Lucas~I.~C.~Medeiros $^{1}$\orcidK{} }
}

\AuthorNames{Júlia da L. A. Silva, Vicente A. de Sousa, Jr., Marcio E. C. Rodrigues, Fred S. R. Pinheiro, Gutembergue S. da Silva, Halysson B. Mendonça, Ricardo Q. de F. H. Silva, João V. L. da Silva, Fernanda E. S. Galdino, Vitor F. C. de Carvalho and Lucas~I.~C.~Medeiros}

\AuthorCitation{\hl{da L. A. Silva, J.;} 
 A. de Sousa, V., Jr.; E. C. Rodrigues, M.; S. R. Pinheiro, F.; S. da Silva, G.; B. Mendonça, H.; Q. de F. H. Silva, R.; V. L. da Silva, J.; E. S. Galdino, F.; F. C. de Carvalho, V.; I.~C.~Medeiros, L.}

\address{%
$^{1}$ \quad Graduate Program in Electrical and Computer Engineering (PPgEEC), Federal University of Rio Grande do \textls[-15]{Norte, Natal 59078-970, Brazil; }
\mbox{\hl{julia.andrade.097@ufrn.edu.br (J.d.L.A.S.); } 
} \mbox{ \textls[-5]{lucas.medeiros.114@ufrn.edu.br (L.I.C.M.)}}\\
$^{2}$ \quad Department of Communications Engineering, Federal University of Rio Grande do Norte, \mbox{Natal 59078-970, Brazil;} marcio.rodrigues@ufrn.br (M.E.C.R.); fred.rossiter@ufrn.br (F.S.R.P.); gutembergue.soares@ufrn.br (G.S.d.S.); fernanda.galdino.075@ufrn.edu.br (F.E.S.G.) \\
$^{3}$ \quad National Telecommunication Agency (ANATEL), Natal 59020-200, 
 Brazil; halysson@anatel.gov.br\\
$^{4}$ \quad\hl{ Department} 
 of Electrical Engineering, Federal University of Rio Grande do Norte, Natal 59078-970, Brazil; ricardo.queiroz.105@ufrn.edu.br (R.Q.d.F.H.S.); vinicius.lira.116@ufrn.edu.br (J.V.L.d.S.); vitor.fassanaro.089@ufrn.edu.br (V.F.C.d.C.)\\}

\corres{Correspondence: vicente.sousa@ufrn.br}

\abstract{It is crucial to monitor the levels of Non-Ionizing Radiation (NIR) to which the general population may be exposed and compare them to the limits defined in the current standards, in view of the rapid rise of communication services and the prospects of a connected society. A high number of people visits shopping malls and since these locations usually have several indoor antennas close to the public, it is therefore a kind of place that must be evaluated. Thus, this work presents measurements of the electric field in a shopping mall located in Natal, Brazil. We proposed a set of six measurement points, following two criteria: places with great the flow of people and the presence of one or more Distributed Antenna System (DAS), co-sited or not with WiFi access points. Results are presented and discussed in terms of the distance to DAS (conditions: near and far) and flow density of people in the mall (scenarios: low and high number of people). The highest peaks of electric field measured were 1.96 and 3.26 V/m, respectively corresponding to 5\% and 8\% of the limits defined by the International Commission on Non-Ionizing Radiation Protection (ICNIRP) and the Brazilian National Telecommunication Agency (ANATEL).}


\begin{document}

\section{Introduction}
\label{sec:introduction}

Non-Ionizing Radiation (NIR) is described as electromagnetic radiation that has sufficient energy to raise the state of electron excitation, without the capacity of ionizing atoms or molecules~\cite{ICNIRP-1998}. There is an international concern about global health with the increase in artificial forms of NIR sources since its effects on the human body could have the potential to be harmful due to both thermal (rise in body temperature) and non-thermal effects (interaction with the human nervous system)~\cite{world2002}. Thermal effects are better understood, and so, usually, exposure limits are defined in terms of the rise in human body temperature as explained in the recommendations and guidelines on this topic. Because of this, many organizations have created guidelines for human exposure limits, for instance the International Commission on Non-Ionizing Radiation Protection (ICNIRP) in a global context and the Brazilian National Telecommunications Agency (ANATEL) in the Brazilian national scenario. 

Despite the benefits of the accelerating rise of communication services and the prospects of an even more connected society, it is extremely important to monitor the radiation levels to which the general population might be exposed and compare them to the current national and international regulations, especially when exposure limits or measurement procedures have changed in some way. Additionally, the analysis of how scientific publications are distributed in the world is important, in order to map and report NIR studies throughout the globe, along the time.

There is a considerable amount of literature addressing the effects of NIR exposure and how the human body reacts to it in different situations. Although most of the achieved results are inconclusive, it is still of interest to monitor these levels, given the rise of sources to which the population may be exposed now and in the future.

The Brazilian agency ANATEL has a series of practical procedures for measuring the necessary parameters to assess human exposure to NIR. The recommended measurement spots for the analysis of the general population's exposure are:

\begin{itemize}
 \item Points aligned with the direction of maximum radiation from the transmitting station;
 \item Areas with a high flow density of people.
\end{itemize}

Traditionally, shopping malls have a cellular Distributed Antenna System (DAS), which is a set of antennas of mobile cellular signals targeting voice and data indoor coverage for connectivity improvement. Currently, such in-building DASs, generally co-sited with WiFi access points, aggregate the signals of several operators to provide coverage and capacity for 4G LTE and 5G NR. While successfully dealing with the problem of poor outdoor to indoor coverage, the shopping mall's in-building wireless system requires antennas close to the end-users.

As a consequence, shopping malls are places that suit the interests of the measurements, because of the huge flow density of people, the large amount of indoor antennas, and their location relative to the end-users.

Thus, this paper presents and discusses NIR measurements taken in a shopping mall with an installed DAS infrastructure. It is organized as follows. Section~\ref{sec:related-works} presents a quantitative and qualitative analysis of related works about NIR and the measurement methodology of already published papers in Web of Science (WoS) and Scopus databases. Section~\ref{sec:methodology} introduces our measurement setup/methodology while presenting and discussing both the ANATEL and the ICNIRP recommendations about NIR human exposure levels. Section~\ref{sec:results} presents and discusses the measurement results of all evaluated spots, highlighting scenarios with a low and a high number of people inside the shopping mall. Finally, Section~\ref{sec:conclusions} summarizes the paramount results and identifies our further investigations.


\section{Related Works}
\label{sec:related-works}

NIR is a multidisciplinary subject, relating technology and engineering to health and encompassing fields such as biology, medicine, and telecommunications. As a consequence, it arouses the interest and concerns of researchers and the population. This section presents some research investigations involving NIR, showing the research community's interest in the last few decades. Many of these studies focus on different aspects of the NIR measurements and the comparison of measured values to the established human exposure limits.


\subsection{Quantitative Analysis}
\label{subsec:quantitative-analysis}

A survey was carried out on the Scopus and Web of Science (WoS) databases to gather scientific work on NIR. The output data were organized into four groups of keywords. For Groups \#1, \#2, and \#3, we searched on the title, abstract, and keywords of the publications, but we restricted the searching to the title for Group \#0 to avoid papers outside the target area. All searches were from 2002 to 2022. The creation of different searching strategies helped us to identify the general number of papers related to NIR, as well as the specific number of contributions focusing on indoor and outdoor measurements. Table~\ref{tab:articles-keywordsgroups} shows the search strategies.

\begin{table}[H]
 \centering
 \caption{Number of papers by keyword groups.}
 \label{tab:articles-keywordsgroups}
	\newcolumntype{C}{>{\centering\arraybackslash}X}
\begin{tabularx}{\textwidth}{ccccc}
	\hline 
	\multicolumn{1}{l}{\footnotesize \textbf{\makecell{Group}}} & {\footnotesize \textbf{\makecell{Keywords}}} & {\footnotesize \textbf{\makecell{Search within}}} &
 \multicolumn{1}{l}{\footnotesize \textbf{\makecell{Scopus}}} &
 \multicolumn{1}{l}{\footnotesize \textbf{\makecell{WoS}}} \\ \hline
 \footnotesize \#0   & \footnotesize ``Non-ionizing'' AND ``Radiation'' & \footnotesize Title & \footnotesize 236 & \footnotesize 177  \\
	\footnotesize \#1   & \footnotesize \makecell{``Non-ionizing'' AND ``Radiation'' AND ``Measurement''} & \footnotesize \makecell{Title, abstract \\ and keywords} & \footnotesize 636 & \footnotesize 480  \\
	\footnotesize \#2   & \footnotesize \#1 AND ``Outdoor'' & \footnotesize \makecell{Title, abstract \\and keywords} & \footnotesize 29 & \footnotesize 35 \\
 \footnotesize \#3   & \footnotesize \#1 AND ``Indoor'' & \footnotesize \makecell{Title, abstract \\and keywords} & \footnotesize 35 & \footnotesize 31\\
	\hline
	\end{tabularx}
\end{table}

We double-checked the Group \#0 results to guarantee that the papers were exactly related to the effects of NIR, its measurement, or its application. By reading the title, the abstract, and the conclusion of the 50 most-cited papers in this group in Scopus (the database with the largest number of papers found), a classification was performed in terms of:

\begin{itemize}
 \item \textbf{Approach:} measurements/laboratory experiments, reviews and associated discussions, surveys, or legislation;
 \item \textbf{Conclusion}: pessimistic, neutral, or optimistic. 
\end{itemize}

Regarding the conclusion classification, the three options were related to the possible harmful effects of NIR to human health: pessimistic papers concluded the existence of such effects, whereas the optimistic ones concluded the non-existence of the effects, and Neutral papers were inconclusive.

\begin{figure}[H]
 \includegraphics[width=0.8\linewidth]{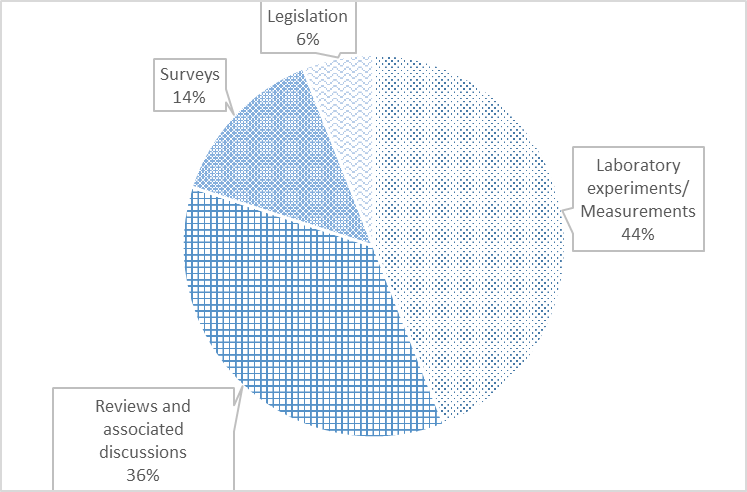}
 \caption{Number of papers by approach.}
 \label{fig:paper-classification}
\end{figure}

\begin{figure}[H]
 
 \includegraphics[width=0.9\linewidth]{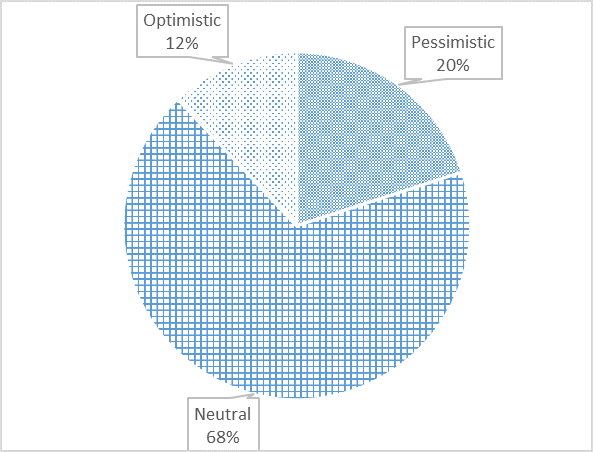}
 \caption{Number of papers by conclusion.}
 \label{fig:paper-conclusion}
\end{figure}

Results for approach classification are shown in Figure~\ref{fig:paper-classification}, whereas the ones for conclusion classification are presented in Figure~\ref{fig:paper-conclusion}.

From Figure~\ref{fig:paper-classification}, we concluded that most publications deal with laboratory experiments or with measurements. Laboratory experiments generally encompass experiments based on exposure to NIR in animals and plants. In the background, there are papers that address reviews or subjective discussions. Regarding the conclusion classification, Figure~\ref{fig:paper-conclusion} shows that the majority of authors advocate a neutral stance on the matter.

The search based on Group \#1 returned several publications extending to areas such as biology and medicine. Then, in order to gather only publications about measuring the general population's exposure to NIR in indoor or outdoor environments due to telecommunication services or home electric appliances, for instance, we refined the searching strategy, resulting in Groups \#2 and \#3.
 
The first publications of Group \#2 were registered in the databases dating from 2006 (Scopus and WoS). However, for Group \#3, the first publications were registered earlier, in 2003 (WoS) and 2005 (Scopus).

\begin{figure}[H]

 \includegraphics[width=0.9\linewidth]{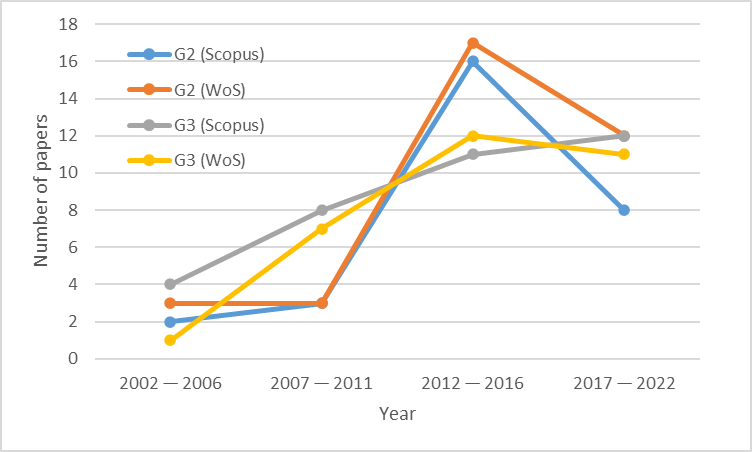}
 \caption{\hl{Number} 
 of papers by year.}
 \label{fig:paper-year}
\end{figure}

Figure~\ref{fig:paper-year} presents a more detailed drawing of the scientific development of indoor and outdoor measurements of the general population's exposure to NIR over time. It shows the number of publications in the last 20 years for Group \#2 (G2) and Group \#3 (G3) for both databases. There was an increase in the number of publications for both groups. However, Group \#3 showed a more persistent scientific interest than Group \#2.

\FloatBarrier

\begin{figure}[H]

 \includegraphics[width=0.9\linewidth]{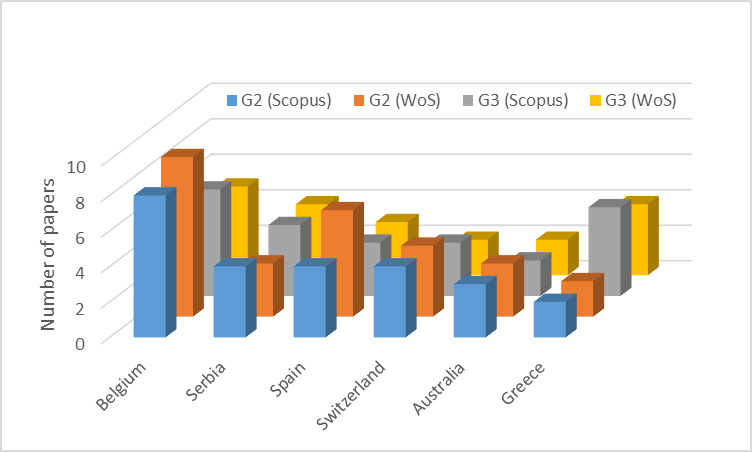}
 \caption{Number of papers by country.}
 \label{fig:paper-country}
\end{figure}

The next step in the quantitative analysis was the search for publications by country to identify the places most committed to discussing the NIR subject. Figure~\ref{fig:paper-country} shows the countries having the highest number of publications on NIR regarding Groups \#2 and \#3. It is worth mentioning that, in many countries, in-field measurements of EM values are the responsibility of specific competent authorities and are not published as scientific papers, but as technical reports not found in the databases used to build this figure. Belgium stands out in the number of publications, ahead of all other countries for both groups. Another interesting fact is that only for Greece, a significant difference between the number of papers in Groups \#2 and \#3 was verified, the latter including indoor studies. 

\subsection{Qualitative Analysis}
\label{subsec:quanlitative-analysis}

This section presents a qualitative analysis of the works found in Group \#3, the searching group more related to this work. Among the first works recorded, a considerable amount carried out measurements of electric and magnetic fields due to the operation of internal power distribution substations at 50~Hz~\cite{safigianni2005,safigianni2007,keikko2009}. We noticed that the indoor environment is the focus of investigation in two main scenarios: the emission of electrical appliances at low~\cite{calvente2014,ghanbari2022,halgamuge2018} and intermediate frequencies~\cite{aerts2017}, as well as the emission of Radio Frequencies (RFs), such as from GSM/UMTS~\cite{koprivica2014} and 4G/LTE~\cite{lunca2014} base stations. In the telecommunications context, there are also narrowband measurements involving different telecommunications services for different environments such as buildings, schools, hospitals, and others~\cite{lahham2015}.

Few studies were found on NIR measurements in shopping malls~\cite{ramirez2021,onishi2021,engiz2017}. In~\cite{ramirez2021}, the measurements were focused on WiFi on both the 2400–2500 MHz and 5150–5850 MHz Industrial Scientific and Medical (ISM) bands. NIR was quantified using the personal display of volunteers for 24 h while performing routine daily tasks in different environments, including shopping malls~\cite{ramirez2021}. The investigations in~\cite{onishi2021} assessed radio frequency exposure levels of outdoor cellular base stations and inside an underground mall in Japan~\cite{onishi2021}. Finally, the authors of \cite{engiz2017} provided broadband measurements in a shopping center, during 24 h a day, for a week, at a fixed point in the place~\cite{engiz2017}. Table~\ref{tab:papers} summarizes the main points of the cited publications, highlighting the comparison to our proposed study in terms of:

\begin{itemize}
 \item \hl{Measurement time follows ICNIRP recommendations: 
} we followed a measurement time of 6 min, according to \cite{ICNIRP-1998,ANATEL_ATO_458}; 
 \item {\hl{Measurement of unperturbed values follows ICNIRP recommendations:}} we isolated the measurement setup to prevent perturbation of the electromagnetic fields due to the flow of people nearby, and we used a wooden tripod to avoid reflective materials near the measuring equipment, according to~\cite{ICNIRP-1998,ANATEL_ATO_458};
 \item {\hl{Study of the influence of the number of people in the shopping mall on NIR levels:}} we analyzed all measurement spots for two scenarios of flow density of people (low and high number of people); we claim that a high number of people potentially loads the mobile networks of the shopping mall, leading to a higher EM level; 
 \item {\hl{Study of the influence of distance to the antenna on NIR levels:}} we performed NIR measurements at two different distances from the antenna, as stated in our Methodology.
\end{itemize}

\begin{table}[H]
 \centering
 \caption{Summary of key research papers on NIR measurements in shopping malls.}
 \label{tab:papers}
	\begin{tabular}{ccccc}
	\hline 
	\multicolumn{1}{l}{\footnotesize \textbf{\makecell{Reference}}} &
 \multicolumn{1}{l}{\footnotesize \textbf{\makecell{Measurement Time \\ Follows ICNIRP \\ Recommendations}}} &
 \multicolumn{1}{l}{\footnotesize \textbf{\makecell{Measurement of \\ Unperturbed Values \\ Follows ICNIRP \\ Recommendations}}} &
 \multicolumn{1}{l}{\footnotesize \textbf{\makecell{Study of Influence \\ of the Number of \\ People in the Shopping \\ Mall on NIR Levels}}} & {\footnotesize 
	\textbf{\makecell{Study of Influence \\ of Distance to \\ the Antenna \\ on NIR Levels}}} \\ \hline
	\footnotesize \cite{ramirez2021}   & \footnotesize \checkmark & \footnotesize x & \footnotesize x & \footnotesize x \\
	\footnotesize \cite{onishi2021} & \footnotesize x & \footnotesize \checkmark & \footnotesize x & \footnotesize x \\
 \footnotesize \cite{engiz2017} & \footnotesize \checkmark & \footnotesize x & \footnotesize \checkmark & \footnotesize x \\
 \footnotesize \makecell{Proposed \\Study} & \footnotesize \checkmark & \footnotesize \checkmark & \footnotesize \checkmark & \footnotesize \checkmark \\
	\hline
	\end{tabular}
\end{table}



Therefore, motivated by the existence of few studies that investigated the levels of exposure in shopping malls and knowing the importance of this study, this paper provides NIR measurements carried out at a shopping mall in Natal, RN, Brazil. Our key contributions are:

\begin{itemize}
 \item Qualitative and quantitative analyses of the state-of-the-art in NIR investigations regarding indoor and outdoor measurements;
 \item Unprecedented measurements to verify compliance with local and global standards of human exposure to NIR from an indoor cellular DAS with nearby WiFi access point installations inside a shopping mall;
 \item Investigation of the dependence between the NIR level and the number of people in the shopping mall, as discussed in~\cite{engiz2017};
 \item Investigation of the relation of NIR levels and the distance to DAS antennas.
\end{itemize}

\section{Methodology}
\label{sec:methodology}

In this section, we present our investigation methodology in terms of the procedure and setup, as well as the place and schedule of measurements. All the measurement rounds were based on recommendations and guidelines defined by ANATEL and ICNIRP.

Brazilian NIR recommendations are defined by the Brazilian federal law 11.934/2009~\cite{Lei11934} and the act 458/2019~\cite{ANATEL_ATO_458}, which sets limits to human exposure to NIR based on ICNIRP guidelines. Exposure limits for the general population are illustrated in Table \ref{tab:limits}. They are considered unperturbed values, i.e., to be compared to measurements in the absence of potentially exposed individuals nearby and without the introduction of absorbing objects or wave reflectors during the measurement process~\cite{ANATEL_ATO_458}. The $f$ value has the same units as indicated in the Radio Frequency Range column~\cite{ANATEL_ATO_458}, and the lower frequency is used to consider a conservative (severe) case.


\begin{table}[H]
 \centering
 \caption{Exposure limits for the general population (unperturbed effective values). Adapted from~\cite{ANATEL_ATO_458}.}
 \label{tab:limits}
\newcolumntype{C}{>{\centering\arraybackslash}X}
\begin{tabularx}{\textwidth}{CCC}
	\hline 
	\multicolumn{1}{l}{\footnotesize \textbf{\makecell{Radio Frequency Range}}} & {\footnotesize \textbf{\makecell{Field Intensity, \\ E (V/m)}}} & {\footnotesize \textbf{\makecell{Equivalent Plane Wave Power \\ Density, S$_{eq}$ (W/m$^2$)}}} \\\hline
	\footnotesize 10 MHz to 400 MHz & \footnotesize $28$ & \footnotesize $2$ \\
	\footnotesize 400 MHz to 2000 MHz & \footnotesize $1.375\sqrt{f}$ & \footnotesize $f/200$ \\
	\footnotesize 2 GHz to 300 GHz & \footnotesize 61 & \footnotesize 10 \\ 
 \hline \\
	\end{tabularx}
\end{table}

As we performed a broadband measurement covering the frequency range from 100~kHz to 6 GHz, it is important to identify the frequencies used by the services in the shopping mall to define the exposure limits. According to ANATEL [18], the spectrum allocated to mobile network operators in Brazil spans from 880 MHz to 2600 MHz. As mentioned before, the shopping mall also has some WiFi access points at 2.4 and 5 GHz. Thus, according to the equation in Table 3, the calculated exposure limit used in our analysis is 40.79 V/m \mbox{(f = 880 MHz)}, representing the most severe case (the lowest value for our target frequency range).


According to \cite{ANATEL_ATO_458}, we measured electric fields considering:

\begin{itemize}
 \item Measurements at points of maximum radiation from the transmitting station being evaluated;
 \item Measurements in areas of a great flow density of people; 
 \item For radio frequencies between 100 kHz and 10~GHz, the time average of 6 min of measurements.
\end{itemize}


\subsection{Measurement Procedure} 
\label{subsec: positioning}

The DAS antennas installed at the shopping mall are the I-ATO1-800/2500 by Radio Frequency Systems (RFS)~\cite{antenna}. In order to confirm that our electric field measurements (V/m) were made in the far field (Fraunhofer) region from the transmitter, we computed the distance from which this region is usually considered to be valid, using Equation (\ref{eqn:farfield}):

\begin{equation}
\label{eqn:farfield}
\frac{2D^2}{\lambda},
\end{equation} \textls[-20]{where $D$ is the largest antenna dimension and $\lambda$ is the operational wavelength. Thus, for the worst case in the target frequency spectrum, we have} $2 \cdot 0.067^2 / \left[ 3 \cdot 10^8 / (2600 \cdot 10^6) \right] = 7.78 \textrm{ cm}$.


As the DAS antennas are positioned in the ceiling of the shopping mall (equivalent height of two residential floors), every possible measurement place at the floor is located in the antenna's far field, since the probe height defined in~\cite{ANATEL_ATO_458} is 1.7 m from the ground. In practice, there are more antenna models installed, but all of them are similar to the I-ATO1-800/2500, basically leading to the same conclusion in terms of far field region distance.

Taking into account the recommendations in~\cite{ANATEL_ATO_458}, the selected point for measurements would consider the distance from the emitter and its radiation diagram. Then, after some preliminary tests, we defined two measurement conditions, the near and the far one. Care must be taken not to confuse the terms \textit{near} and \textit{far} related to antenna distance with near and far field conditions. As mentioned, all measurements were taken at the far field from the transmitter. Considering the origin as the projection of the antenna on the ground, the near measurement point was set at three steps away from the origin, and the far measurement point was set at ten steps from the origin. Therefore, once having defined a relevant azimuth line, near and far measurements were performed in the two opposite orientations along the line: for one orientation, we named it Round 1 measurements and for the opposite one, Round 2~measurements. Thus, there are four measurements for each relevant selected place. Every four points for the same place are labeled with the same number, as depicted in Figure~\ref{fig:setup}.

\subsection{Measurement Setup}
\label{subsec:setup}

The measurement setup includes the following list of equipment: 

\begin{itemize}
 \item Narda Broadband Field Meter NBM-520~\cite{manualnarda};
 \item NBM E-Field-Probe EF 0691 (100 kHz--6 GHz);
 \item Wooden tripod.
\end{itemize}

\begin{figure}[H]

 \includegraphics[width=\linewidth]{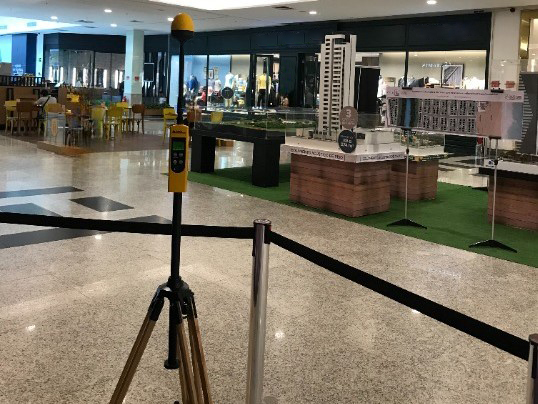}
  \caption{Photo of measurement setup.}
 \label{fig:setup_real}
\end{figure}

In addition, an area isolation strip was used to ensure public distancing within a radius of 2 m from the measuring probe, as defined in~\cite{ANATEL_ATO_458}. Figure~\ref{fig:setup_real} shows a photo of one measurement point.

\begin{figure}[H]

 \includegraphics[width=\linewidth]{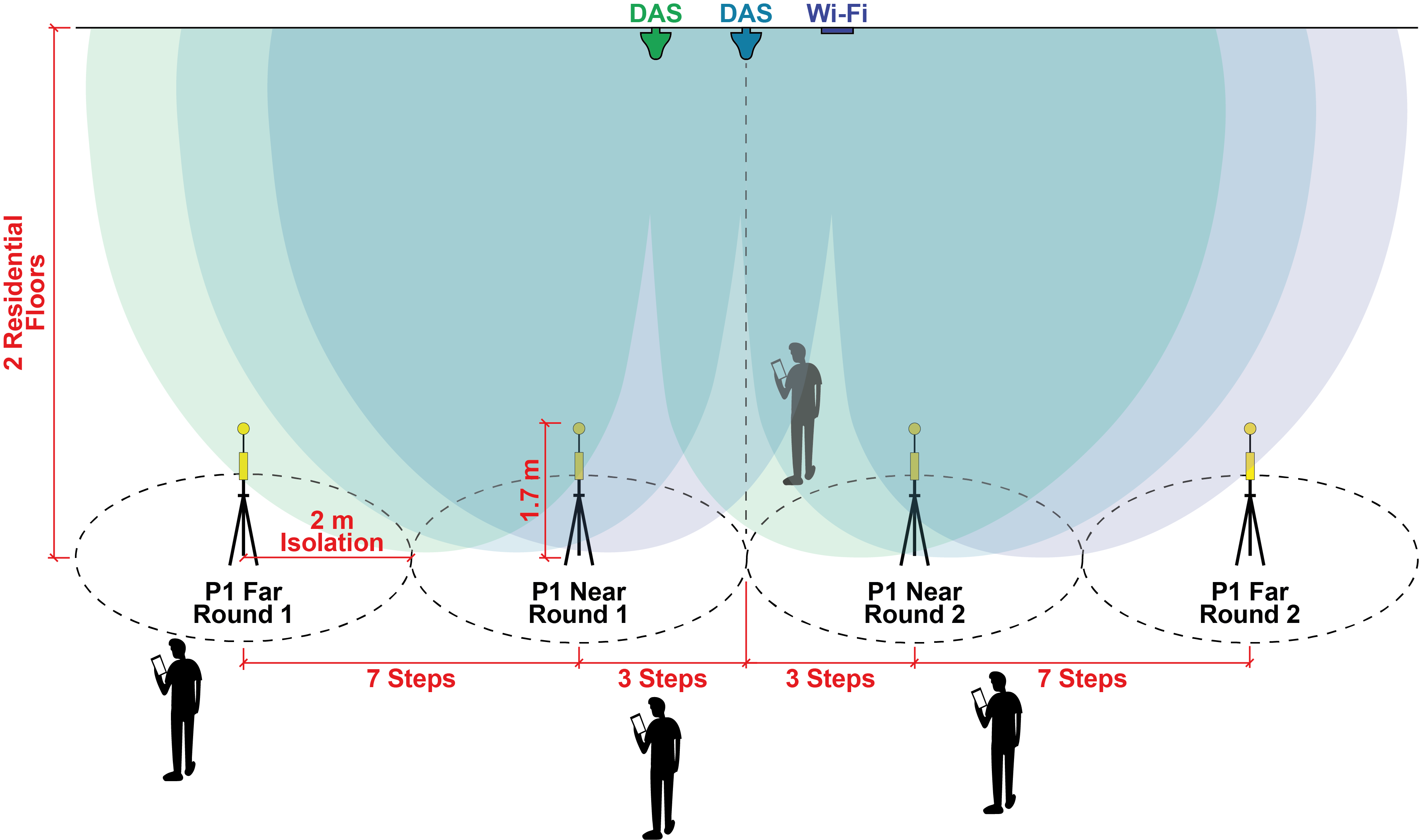}
  \caption{Illustration of the measurement setup for a selected place, including the isolation area and the two distance conditions.}
 \label{fig:setup}
\end{figure}

Figure~\ref{fig:setup} illustrates our measurement setup, including the isolation area and the two distance conditions, near and far. The drawing in this figure illustrates the four measuring points related to the central DAS antenna plotted in the ceiling.

\subsection{Measurement Places' Selection}
\label{subsec:places}

All measurements were accomplished at Natal Shopping, located in Natal, a city in the northeast region of Brazil. The shopping building has three floors composed of long and wide corridors with stores, food courts, and some entertainment areas. It also has a garage building.

All the measurement places were defined taking into account two important requirements:

\begin{itemize}
 \item Presence of one or more DASs, co-sited or not with WiFi access points;
 \item Great flow of people.
\end{itemize}

Thus, a total of six places were established, three places per floor. As stated in Section~\ref{subsec:setup} and illustrated in Figure~\ref{fig:setup}, each selected measurement place gave rise to four measurement points having the number of the place and an indication of its condition category (distance to the DAS antenna) and of its round.

\begin{figure}[H]
 \centering
 \begin{subfigure}{0.3\textwidth}
  \centering
  \includegraphics[width=\textwidth]{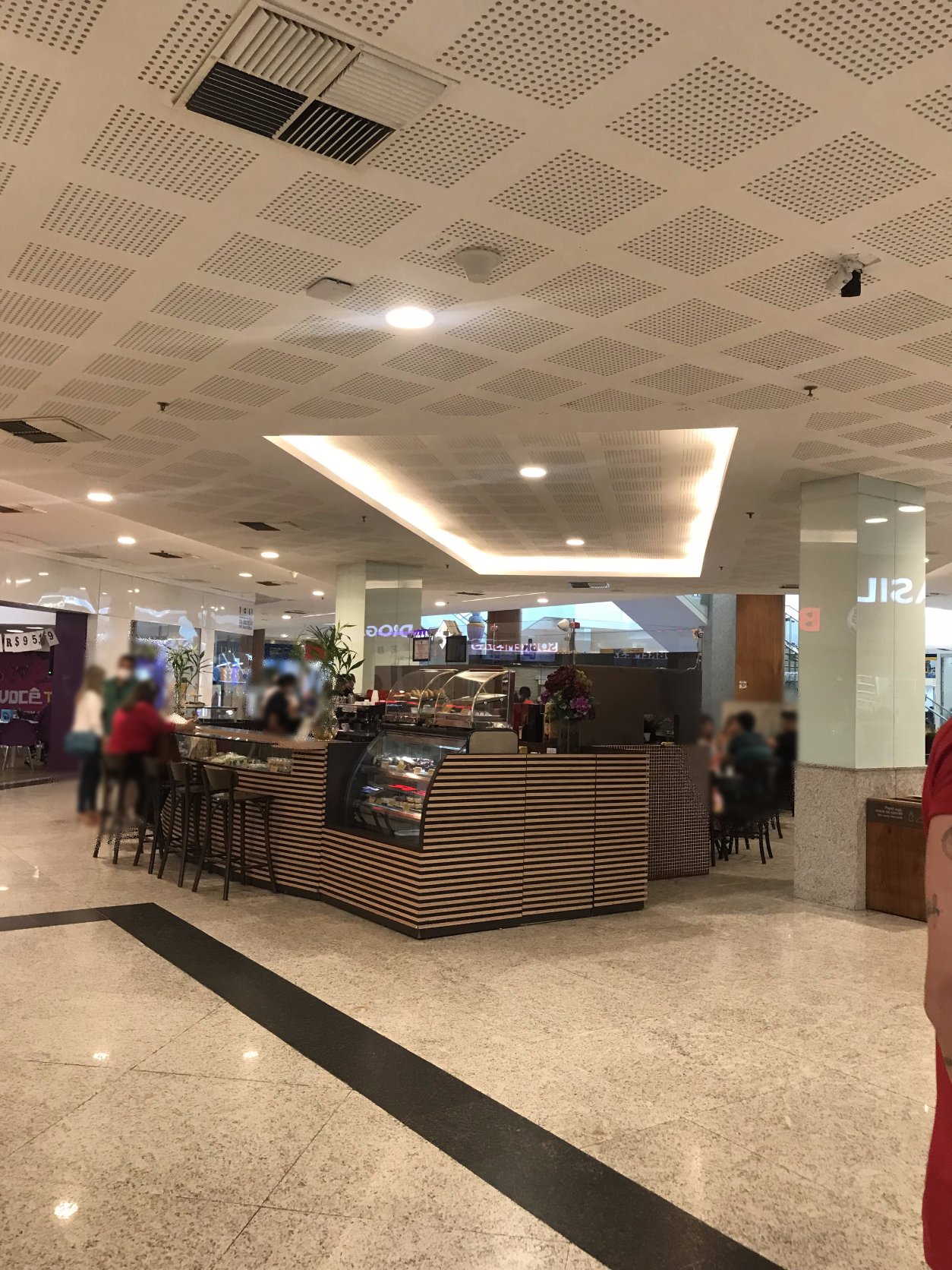}
  \caption{}
  \label{fig:p1l1}
 \end{subfigure}
 \hfill
 \begin{subfigure}{0.3\textwidth}
  \centering
  \includegraphics[width=\textwidth]{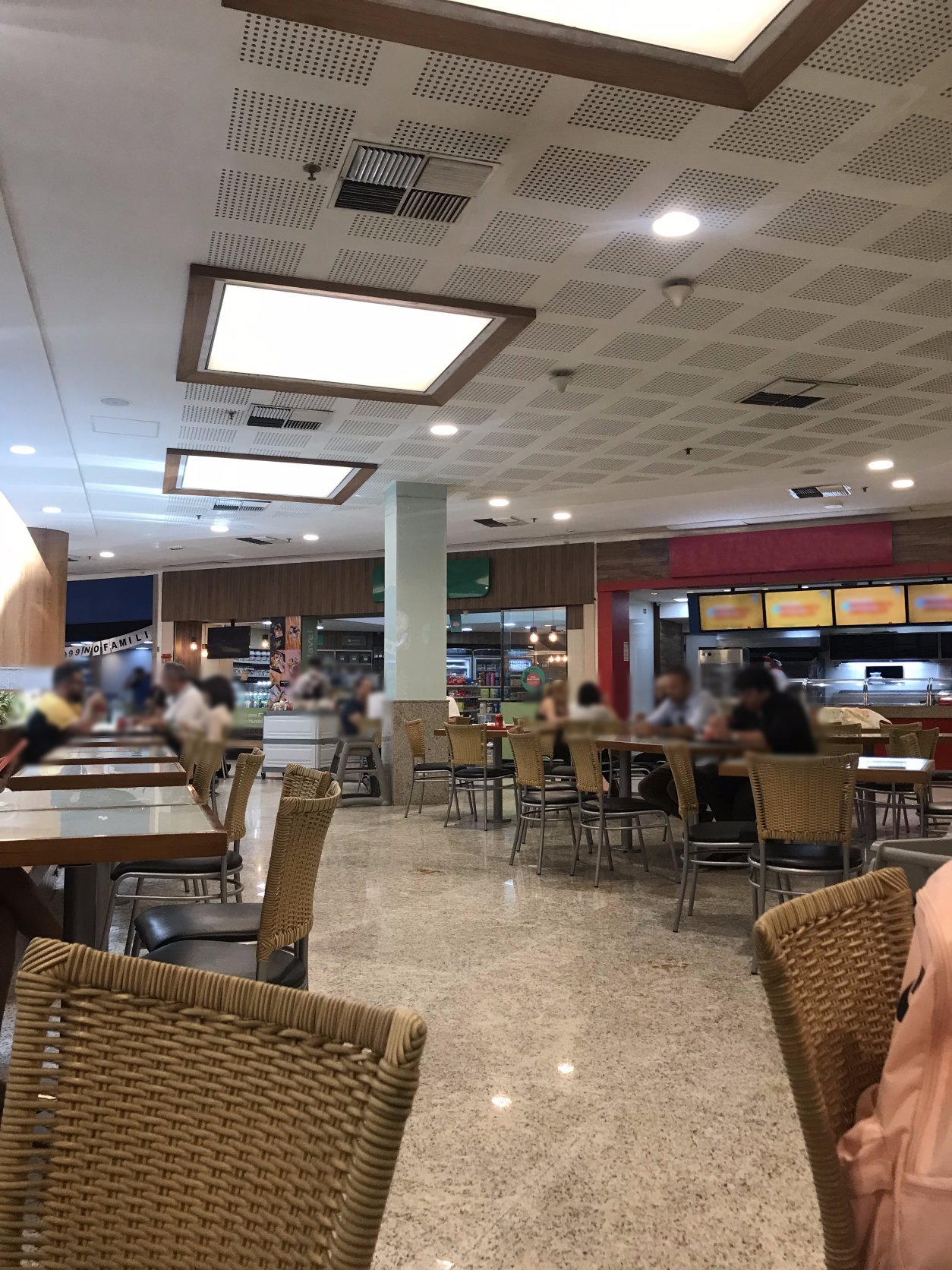}
  \caption{}
  \label{fig:p2l1}
 \end{subfigure}
 \hfill
 \begin{subfigure}{0.3\textwidth}
  \centering
  \includegraphics[width=\textwidth]{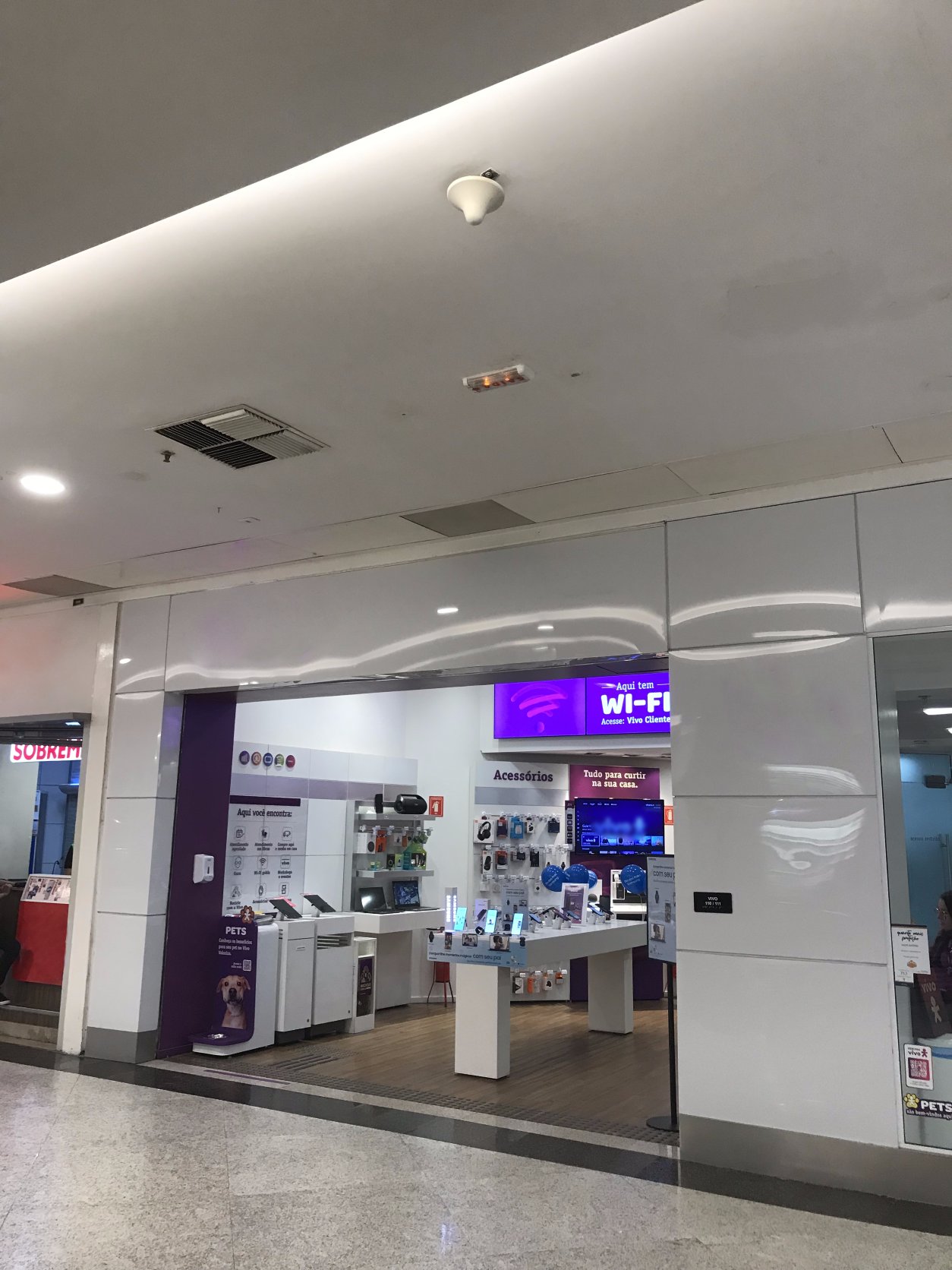}
  \caption{}
  \label{fig:p3l1}
 \end{subfigure}
  \caption{Sample view of selected places at L1 floor. (\textbf{a}) \hl{P1L1;} 
 (\textbf{b}) P2L1; (\textbf{c}) P3L1.}
  \label{fig:points-L1}
\end{figure}

\begin{figure}[H]
 \centering
 \begin{subfigure}{0.3\textwidth}
  \centering
  \includegraphics[width=\textwidth]{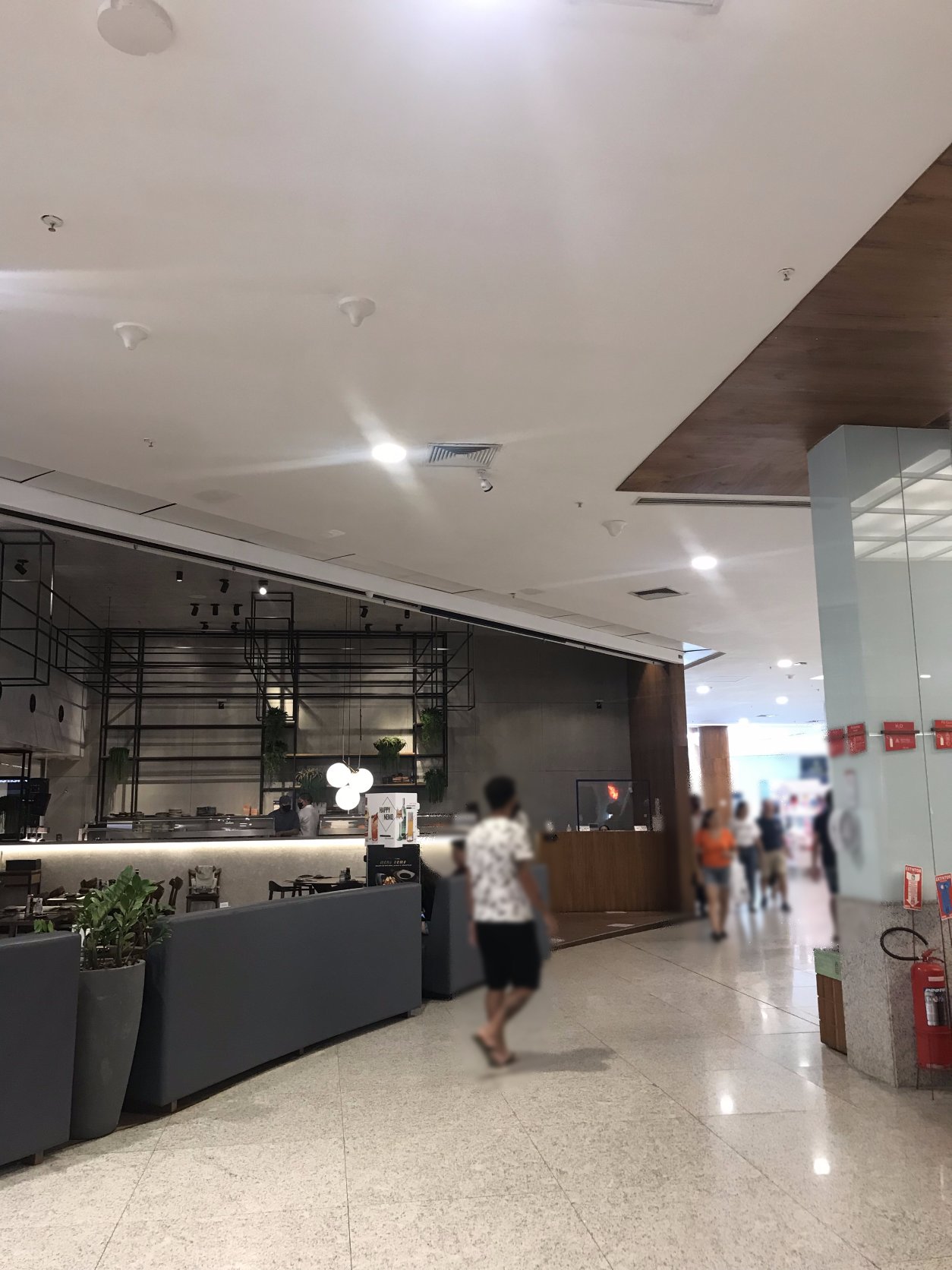}
  \caption{}
  \label{fig:p1l2}
 \end{subfigure}
 \hfill
 \begin{subfigure}{0.3\textwidth}
  \centering
  \includegraphics[width=\textwidth]{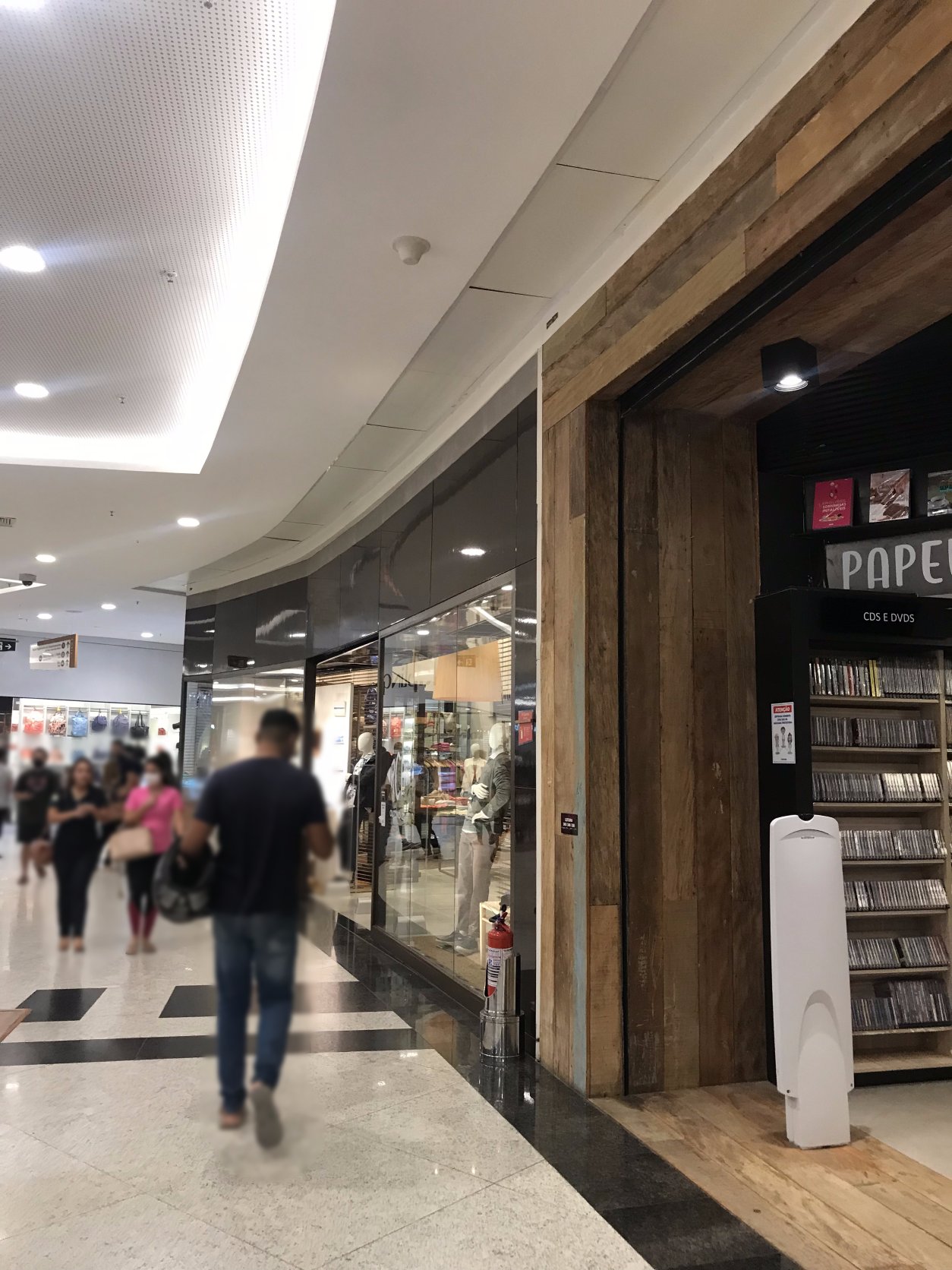}
  \caption{}
  \label{fig:p2l2}
 \end{subfigure}
 \hfill
 \begin{subfigure}{0.3\textwidth}
  \centering
  \includegraphics[width=\textwidth]{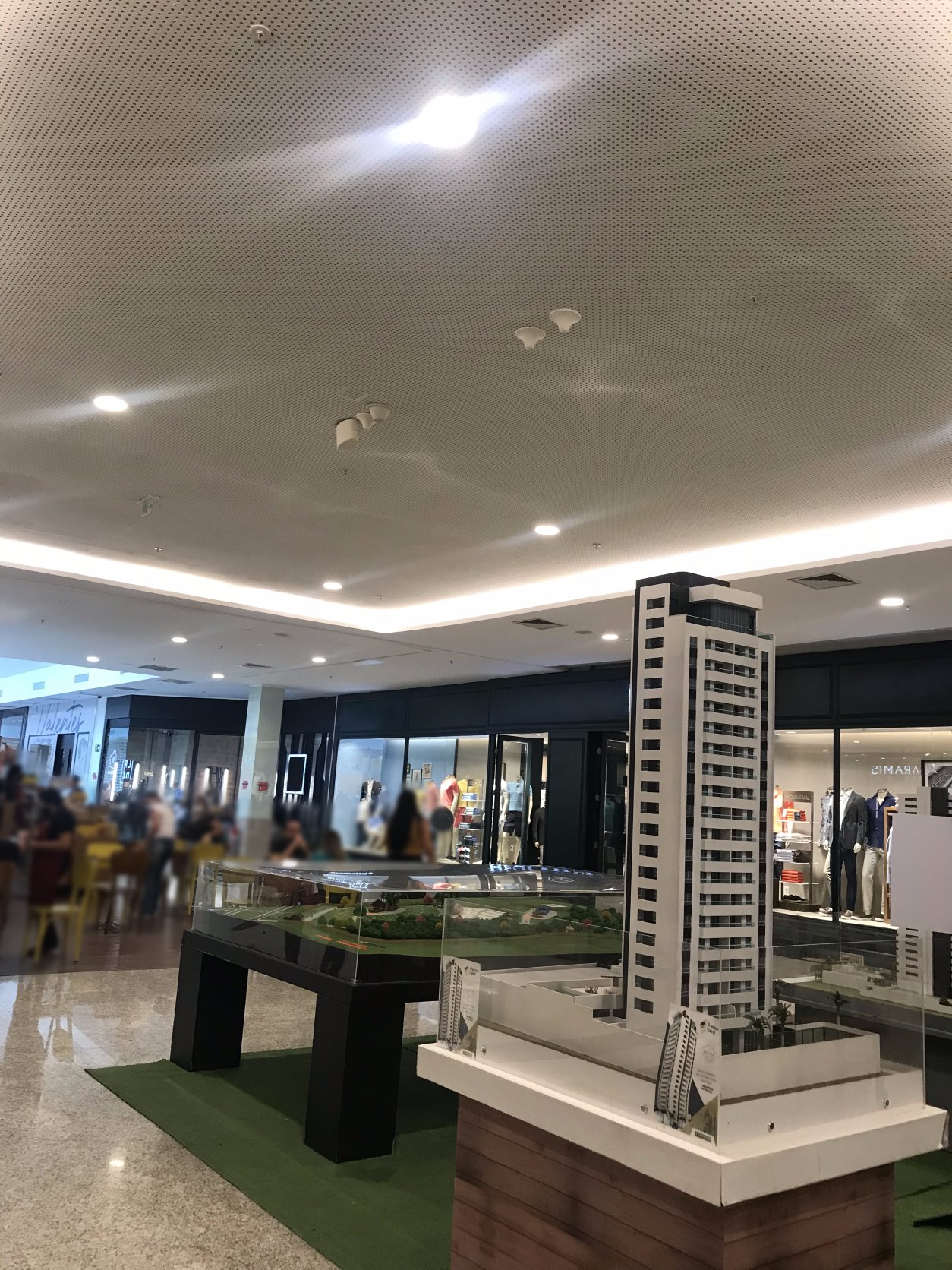}
  \caption{}
  \label{fig:p3l2}
 \end{subfigure}
  \caption{\hl{Sample} view of selected places at L2 floor. (\textbf{a}) {P1L2;} (\textbf{b}) P2L2; (\textbf{c}) P3L2.}
  \label{fig:points-L2}
\end{figure}

In each floor, measurement places were numbered 1 to 3. Suffix L1 or L2 designates places on the first or second floor, respectively. On the first floor, the selected places were located near or inside the main food court. Two places on the second floor (P2L2 and P3L2) were located in the commercial area, and the third one (P1L2) was located next to restaurants. Figures~\ref{fig:points-L1} and \ref{fig:points-L2} present a sample view of the selected places for the first and second floors, respectively. 

\subsection{Measurement Schedule Definition}
\label{subsec:schedule}

Visiting the website of Natal Shopping on Google, we used information from the Popular Times~\cite{googlepopular} to verify the flow density of people categorized by days of the week and time. Google metrics showed a low flow density of people at 10~a.m. (mall opening time) and a high flow density at 4 p.m., especially on Fridays. Thus, we had two measurement scenarios: low- and high-flow density of people.

Therefore, our measurement campaigns were conducted on two different days of the week and at two different times in order to include the two scenarios with the following schedule:

 \begin{itemize}
  \item Tuesday morning (10 a.m.--12 noon): low number of people;
  \item Friday afternoon (4 p.m.--6 p.m.): high number of people.
 \end{itemize}


Measurements were performed along two weeks: the first one dedicated to the second floor and the second one dedicated to the first floor.

Following the measurement conditions defined in Section~\ref{subsec: positioning} this section presents the schedule, which made it possible to define two desired scenarios. Then, we have two \hl{conditions} 
 ({\hl{near} 
} and {\hl{far}}), while we considered two {\hl{scenarios}} of the flow of people in the mall ({\hl{low}} and {\hl{high}} number of people). To make possible the comparison, selected measurement points were exactly the same for both scenarios. The next section presents and discusses our results.

\subsection{{Assumptions and Limitations}}

{Although we followed the ANATEL and ICINIRP recommendations and guidelines to build our measurement setup, some assumptions must be highlighted.}

{An important assumption is related to the RF characterization of the EM sources. As we focused on already installed telecom infrastructure, we did not have access to all deployment projects, as well as detailed antenna information. However, it must be remarked that, even considering the access to every RF emitter's datasheet, the determination of maximum exposure points from them would not be assured: the interaction between the antenna and ceiling features usually generates some distortion in the radiation diagram. Therefore, the most accurate way to find peak exposure levels is via a survey in the vicinity of the theoretically worst-case points. We performed some preliminary searching before the definitions of the two measurement conditions, the near and the far one. Thus, regarding the broad antenna elevation diagram, all defined measurement points were in the maximum radiation region, even though we could not precisely guarantee the maximum radiation point.}

{We also considered that the exposure condition is related to the density of people in the mall. Although our paper provides and discusses results for two scenarios of the flow density of people, the {\hl{high-flow density of people}} scenario defined in Section~\ref{subsec:schedule} could not ensure a worst-case scenario because we cannot guarantee that all EM sources are using their maximum transmission power. Since we focused on realistic scenarios, the only way to obtain different system loads, consequently different EM source loads, was to take measurements with a high- and low-flow density of people in the mall. Indeed, this is the methodology recommended by ICNIRP and ANATEL, but we could consider EM levels from a dedicated base station in our further investigations, perhaps a future cooperation among our university, the ANATEL agency, and a telecom operator partner.}



\section{Results and Discussions}
\label{sec:results}

Table~\ref{tab:peakvalues} shows the peak values found for each target point selected measurement place, indicating the conditions, scenarios, and Exposure Ratio (ER), the ratio between the measured value and the limit, expressed as a percentage for two exposure limits:

\begin{itemize}
 \item Suitable situation of the accomplished campaigns (measurements of DAS signals that are down-limited to f~=~800~MHz); 
 \item Very conservative (pessimistic) situation (including FM broadcast frequencies from f~=~88~MHz).
\end{itemize}

 Our results indicate that the shopping mall's NIR levels are below the regulation limit. The highest average and peak electric fields found were 1.96 and 3.26 V/m, respectively, referring to the P3L2 place. Despite this, the values correspond to approximately 5\% and 8\% of the limit defined by ICNIRP and ANATEL, respectively, using f~=~880~MHz for calculating the limit. The highest peak electric fields found were also below that measured in~\cite{engiz2017}, which was 7.88~V/m. However, the maximum average found in~\cite{engiz2017} was lower, with a value of 0.87~V/m.
 
 Trying to be stricter with the limit, we considered the FM band with f~=~88~MHz, finding a limit of 28~V/m, as indicated in Table~\ref{tab:limits}. Naturally, the ER increases when using the lower FM frequency, yet far from the exposure limit.


\begin{table}[H]
 \centering
 \caption{ Peak value and ER for each selected measurement place.}
 \label{tab:peakvalues}
	\newcolumntype{C}{>{\centering\arraybackslash}X}
\begin{tabularx}{\textwidth}{cCCCCC}
	\hline 
	\multicolumn{1}{l}{\footnotesize \textbf{\makecell{Place}}} & {\footnotesize \textbf{\makecell{Point Condition \\ (Distance)}}} &
 \multicolumn{1}{l}{\footnotesize \textbf{\makecell{Scenario \\ (Number of People)}}} &
 \multicolumn{1}{l}{\footnotesize \textbf{\makecell{Peak Electric \\ Field (V/m)}}} &
 \multicolumn{1}{l}{\footnotesize \textbf{\makecell{ER (\%) \\ (f = 880~MHz)}}} & \multicolumn{1}{l}{\footnotesize \textbf{\makecell{ER (\%) \\ (f = 88~MHz)}}} \\ \hline
 \footnotesize P1L1 & \footnotesize Far & \footnotesize High & \footnotesize 1.68 & \footnotesize 4.12 & \footnotesize 6.00 \\
	\footnotesize P2L1 & \footnotesize Near & \footnotesize High & \footnotesize 2.90 & \footnotesize 7.11 & \footnotesize 10.35 \\
 \footnotesize P3L1 & \footnotesize Far & \footnotesize High & \footnotesize 1.68 & \footnotesize 4.12 & \footnotesize 6.00 \\
 \footnotesize P1L2 & \footnotesize Near & \footnotesize High & \footnotesize 2.69 & \footnotesize 6.59 & \footnotesize 9.60 \\
 \footnotesize P2L2 & \footnotesize Far & \footnotesize Low & \footnotesize 1.51 & \footnotesize 3.70 & \footnotesize 5.39 \\
 \footnotesize P3L2 & \footnotesize Near & \footnotesize Low & \footnotesize 3.26 & \footnotesize 7.99 & \footnotesize 11.64 \\ \hline 
	\end{tabularx}
\end{table}

Analyzing the condition category (distance to DAS antennas), we observed higher NIR peak values for the near condition. The elevation pattern of the radiation diagram plays a role in this analysis, and so, we considered that its broad half-power beamwidth, path loss, and propagation conditions are responsible for the lower levels at far distances. 

On the other side, we expected a higher usage of mobile phones as more people are in the shopping mall, leading to higher NIR values. However, it is not possible to identify a strong relation between the scenario (number of people) and the peak values.

\begin{figure}[H]

 \includegraphics[width=0.9\linewidth]{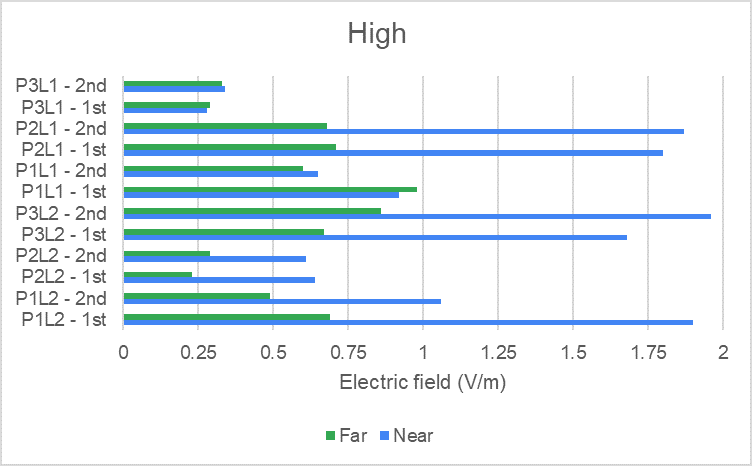}
  \caption{\hl{Average} 
 electric field in the high number of people scenario.}
 \label{fig:high-movement}
\end{figure}

\begin{figure}[H]

 \includegraphics[width=0.9\linewidth]{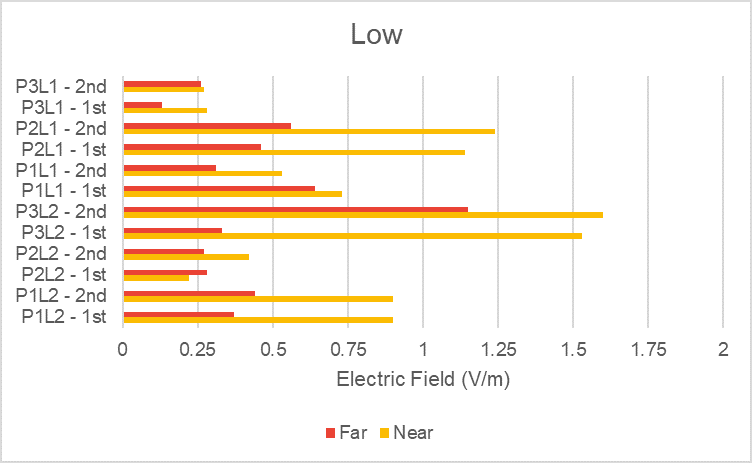}
  \caption{\hl{Average} electric field in the low number of people scenario.}
 \label{fig:low-movement}
\end{figure}

To analyze the NIR average results for the condition (distance to DAS antennas), we grouped the measured data with the same scenario and plot them separately in Figures~\ref{fig:high-movement}~and~\ref{fig:low-movement} for a high and low number of people, respectively. The figures present the average NIR values for the two measurement rounds taken. Rounds are identified with suffixes 1st and 2nd, for each one of the six defined measurement places. Then, considering the total number of places, rounds, and conditions (near and far), the figures present a total of 24 measurement points.

\FloatBarrier

Regarding the relation between the condition category (distance to DAS antennas) and the measured electric field values, we reached the same conclusion as the one presented for the peak electric field values in Table~\ref{tab:peakvalues}: in general, higher average NIR exposure is observed for the near condition.

Some surprising results were noticed, and we highlight the following additional discussions:

\begin{itemize}
 \item The average NIR level for the near condition was lower than for the far condition in one of the measurement rounds for places P1L1 and P3L1 in Figure~\ref{fig:high-movement}, as well as in P2L2 in Figure~\ref{fig:low-movement}. The behavior changed according to the scenario, indicating a possible relation with changes in clutter for a low and high number of people;
 %
 
 %
 \item The points mentioned above also presented the lowest levels of electric field and the smallest variation with the change of distance. Although further investigation must be performed, the small variation with distance suggests peculiar propagation conditions, such as some degree of canalization. In both figures, this is particularly evident for place P3L1, which is a place in a corridor;
 \item On the other hand, places P2L1, P1L2, and P3L2 showed a well-defined behavior regarding the condition category. P2L1 and P1L2 presented more than one DAS antenna and/or WiFi antenna nearby. They are also close to restaurants, where people stay for a long time. P3L2 is the place with the largest number of DAS antennas together, and it is located in the shopping mall corridor. Therefore, as expected, this place showed higher measured levels. Besides having the largest measured values, for these points, the influence of distance becomes more visible. This can possibly be explained by the composition of the radiation diagrams from the multiple DAS antennas positioned at that place.
\end{itemize}

\begin{figure}[H]
 
 \includegraphics[width=\linewidth]{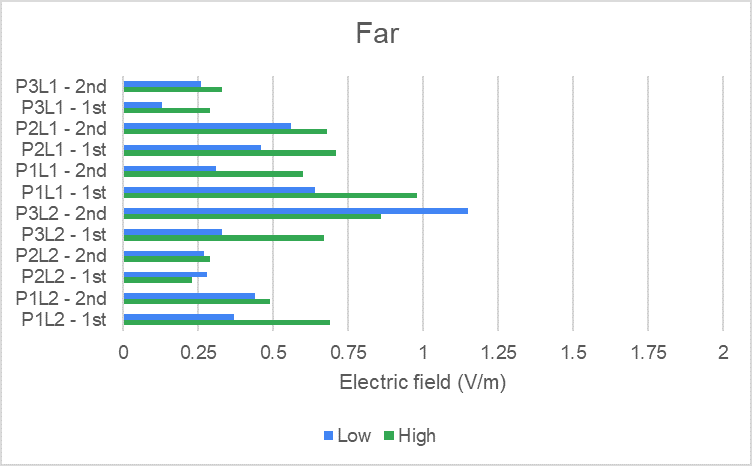}
  \caption{\hl{Average} electric field in far distance condition.}
 \label{fig:far-distanceofantenna}
\end{figure}

\begin{figure}[H]
 
 \includegraphics[width=\linewidth]{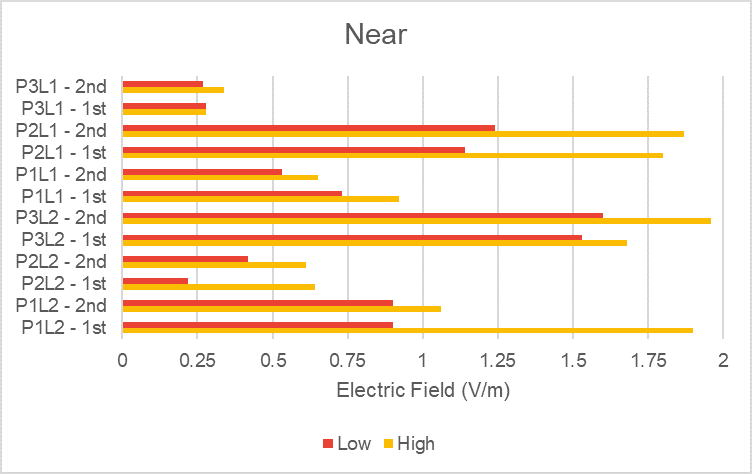}
  \caption{\hl{Average} electric field in near distance condition.}
 \label{fig:near-distanceofantenna}
\end{figure}

Similarly, to analyze the NIR average results of the scenarios (number of people), we grouped measured data with the same condition and plot them separately in Figures~\ref{fig:far-distanceofantenna}~and~\ref{fig:near-distanceofantenna} for the far and near distance of the DAS antenna, respectively. The figures present the average NIR values for the two measurement rounds taken. Rounds are identified in the same way explained for the previous two figures.

Comparing the scenario categories (low and high number of people), the average values showed a well-defined behavior when compared to the peak ones presented in the analysis of Table~\ref{tab:peakvalues}. For the average NIR values, we generally observed a direct relation between the increase in the flow density of people and the increase in NIR exposure. A higher number of people means that the potential users of the indoor mobile network increase, which can lead to a higher measured electric field from user terminals and also possibly from the DAS antennas, since a higher traffic demand may require more transmitters to be activated.

Places P3L2 and P2L2 in Figure~\ref{fig:far-distanceofantenna} present the opposite NIR behavior for the first and the second measurement rounds. Both places had the highest level of electric field for the high number of people scenario, as would be generally expected. However, it happened just for one measurement round, the other one presenting higher levels for the low number of people scenario. Although propagation conditions can be different for each measurement round, especially at the far distance, we also attributed the behavior to the fact that these places are located in corridors, and NIR depends on the number of people using a mobile phone and passing through that corresponding corridor at the time of measurement for each round.

\section{Conclusions}
\label{sec:conclusions}
This work aimed to evaluate the levels of exposure to NIR in shopping malls, considering the few scientific publications in this scenario and being aware of the importance of this study, especially regarding the high circulation of people in these environments and the large number of antennas relatively close to the public.

The measurements carried out showed that the electric fields were significantly below the limits established by ANATEL and ICNIRP. For the target frequency bands in this work, the highest average and peak electric fields were less than 8\% of the limit, with values of 1.96 and 3.26 V/m, respectively. Both results correspond to the P3L2 place, characterized by having the largest number of DAS antennas together.
 
When comparing the change in antenna distance condition for a given number of people scenario, we also verified the strong influence of the number of nearby antennas. Places P2L1, P1L2, and P3L2 have at least two DAS antennas and/or a nearby WiFi antenna, and their measurements showed significant changes between the near and far conditions. Furthermore, these places presented the highest overall levels of exposure to NIR.
 
By grouping measured data corresponding to the same distance condition, we could observe the influence of the number of people on the exposure levels. The highest electric field values were verified at times when the mall was most crowded, as in~\cite{engiz2017}.


Based on the results, the influence of the antenna distance and the concentration of people in the mall on the electric field levels was noted, but in none of the situations did the exposure levels rise above the safe limits established by ICNIRP and ANATEL. However, since there is not a global conclusion, neither from the academic community nor from the World Health Organization (WHO), on the harmful effects of NIR on human health, prevention is always necessary. Thus, besides the general population being oriented to a moderate use of mobile phones, the importance of an adequate project for the distribution of antennas in malls to avoid the intensification of NIR levels caused by the concentration of antennas is verified, mainly in places where the public stays for a long time.

\vspace{6pt} 

\authorcontributions{Conceptualization and methodology: J.d.L.A.S., V.A.d.S.J., M.E.C.R., F.S.R.P., G.S.d.S., and H.B.M.; software and data curation: J.d.L.A.S., R.Q.d.F.H.S., J.V.L.d.S., F.E.S.G., V.F.C.d.C., and L.I.C.M.; validation: J.d.L.A.S., V.A.d.S.J., and M.E.C.R.; formal analysis and investigation: J.d.L.A.S., V.A.d.S.J., M.E.C.R., F.S.R.P., G.S.d.S., and H.B.M.; resources: H.B.M.; writing---original draft preparation: J.d.L.A.S., R.Q.d.F.H.S., J.V.L.d.S., F.E.S.G., V.F.C.d.C., and L.I.C.M.; writing---review, editing and visualization: J.d.L.A.S., V.A.d.S.J., M.E.C.R., F.S.R.P., and G.S.d.S.; supervision: V.A.d.S.J., M.E.C.R., F.S.R.P., G.S.d.S., and H.B.M.. All authors have read and agreed to the published version of the manuscript.}

\funding{\hl{This} 
 study was financed in part by the Coordena\c{c}\~{a}o de Aperfei\c{c}oamento de Pessoal de N\'{i}vel Superior-Brasil (CAPES) Finance Code 001.}

\institutionalreview{\hl{~} Not applicable. 
}

\informedconsent{\hl{~} Not applicable. 
}

\dataavailability{\hl{~} Not applicable. 
}

\acknowledgments{The authors would like to thank the Brazilian National Telecommunications Agency (ANATEL) for the partnership. Special thanks to the operational unit of ANATEL in Natal~city.} 

\conflictsofinterest{The authors declare no conflict of interest.} 


\begin{adjustwidth}{-\extralength}{0cm}

\reftitle{References}

\end{adjustwidth}
\end{document}